\documentclass{kapedbk}

\usepackage{epsfig}
\usepackage{algorithmic}

\let\footnote\savefootnote
\let\footnotetext\savefootnotetext 
\let\footnoterule\savefootnoterule 
\normallatexbib
\startauthorindex

\begin{document}

\newtheorem{lemma}[theorem]{Lemma}
\newtheorem{corollary}[theorem]{Corollary}
\newenvironment{proof}{\noindent\par{\bf Proof: }}{\nopagebreak\rule{1ex}{0.8em}\medskip}

\newcommand{\ddd}{{**}}
\newcommand{\hhh}[1]{\overline{#1}}

\newcommand{\R}{{\bf R}}
\newcommand{\Z}{{\bf Z}}
\newcommand{\PP}{{\cal P}}
\newcommand{\MM}{{\cal M}}
\newcommand{\Mk}{\MM_k}

\newcommand{\mytab}[0]{\hspace{2\labelsep}}
\newcommand{\mt}[0]{\mytab\ }
\newcommand{\mtt}[0]{\mytab\ \mytab\ }
\newcommand{\mttt}[0]{\mytab\ \mytab\ \mytab\ }
\newcommand{\mtttt}[0]{\mytab\ \mytab\ \mytab\ \mytab\ }
\newcommand{\mttttt}[0]{\mytab\ \mytab\ \mytab\ \mytab\ \mytab\ }
\newcommand{\bfbegin}{{\bf begin}}
\newcommand{\bfend}{{\bf end}}
\newcommand{\bffor}{{\bf for}}
\newcommand{\bfdo}{{\bf do}}
\newcommand{\bfendfor}{{\bf endfor}}
\newcommand{\bfendif}{{\bf endif}}
\newcommand{\bfendwhile}{{\bf endwhile}}
\newcommand{\bfreturn}{{\bf return}}
\newcommand{\bfelse}{{\bf else}}
\newcommand{\bfthen}{{\bf then}}
\newcommand{\bfif}{{\bf if}}
\newcommand{\bfwhile}{{\bf while}}
\newcommand{\bfprocedure}{{\bf procedure}}

\articletitle{The Risk Profile Problem for \\ Stock Portfolio 
  Optimization}

\author{Ming-Yang Kao\thanks{Research supported in part by NSF grants
CCR-9531028 and CCR-9974871. Current address: Department of Computer Science,
Northwestern University, Evanston, IL 60201, USA.}}
\affil{
Department of Computer Science 
\\ 
Yale University
\\
New Haven, CT 06520
\\ 
USA}
\email{kao-ming-yang@cs.yale.edu}

\author{Andreas Nolte}
\affil{
Department of Computer Science
\\ 
Yale University
\\ 
New Haven, CT 06520
\\
USA}
\email{nolte@cs.yale.edu}

\and

\author{Stephen R. Tate\thanks{Research supported in part by
Texas Advanced Research Program Grant 1997-003594-019.}}
\affil{
Department of Computer Science
\\ 
University of North Texas 
\\ 
Denton, TX 76203 
\\
USA.}
\email{srt@cs.unt.edu}

\begin{abstract} 
This work initiates research into the problem of determining an
optimal investment strategy for investors with different attitudes
towards the trade-offs of risk and profit.  The probability
distribution of the return values of the stocks that are considered
by the investor are assumed to be known, while the joint distribution
is unknown. The problem is to find the best investment strategy in
order to minimize the probability of losing a certain percentage of the
invested capital based on different attitudes of the investors towards
future outcomes of the stock market.

For portfolios made up of two stocks, this work shows how to  exactly and
quickly solve the problem of finding an optimal portfolio for
aggressive or risk-averse investors, using an algorithm based on a
fast greedy solution to a maximum flow problem.  However, an investor
looking for an average-case guarantee (so is neither aggressive or
risk-averse) must deal with a more difficult problem.  In particular,
it is  $\sharp P$-complete to compute the distribution function associated with the
average-case bound.  
On the positive side, approximate answers can be computed by using
random sampling techniques similar to those for high-dimensional
volume estimation.
When $k>2$ stocks are considered, it is proved that a simple solution based
on the same flow concepts as the 2-stock algorithm would imply that
$P=NP$, so is highly unlikely.  This work gives approximation algorithms for
this case as well as exact algorithms for some important special cases.
\end{abstract}


\newcommand{\port}[1]{\langle{#1}\rangle}

\pagenumbering{arabic}
\section{Introduction} This work  initiates the study of the {\it risk
profile} problem for stock portfolio optimization.  The problem has several
variants depending on a given investor's preference toward the
trade-off between risk and return {\cite{Sharpe:1995:I}}.

In the problem, the investor has a capital, which is normalized to one
dollar.  She considers $k$ different stocks $S_1,\ldots,S_k$ and
wishes to invest some $x_i$ dollars in each stock $S_i$ for a certain
period of time, where $\sum_{i=1}^k x_i=1$ and $x_i \geq 0$ for all
$i$. The vector $\vec{x} = \port{x_i}_{i=1}^k =
\port{x_1,x_2,\ldots,x_k}$ is called a {\it portfolio}.  Let $\PP_k$
be the set of all portfolios for $k$ stocks.  The {\it return} of
$\vec{x}$ is the ratio, expressed as a percentage, of the worth of
this portfolio at the end of the investment period to the initial
investment of one dollar. The {\it return} of stock $S_j$ is the ratio
of its price at the end of the investment period to its initial price,
which is the same as the return of the portfolio $\port{x_i}_{i=1}^k$
with $x_j = 1$ and all the other $x_i = 0$.

\newcommand{\CS}{{\cal S}}

In mathematical finance, stock prices are often assumed to follow
geometric Brownian motions or its variants (e.g., see
\cite{Duffie:1996:DAP,Elliott:1998:MFM,Fouque:2000:DFM,Hull:2000:OFO,Karatzas:1997:LMF,%
Karatzas:1998:MMF,Musiela:1997:MMF}). To complement this conventional
approach with computer science methodologies \cite{Cormen:1990:IA},
we assume that stock prices can move arbitrarily.

Let $\mu$ be a positive real number.  Let $m_1$ and $m_2$ be integers
with $m_1 < m_2$, and let $m= m_2-m_1+1$. Let $\Delta=\{\ell\mu \mid
\ell = m_1,\ldots,m_2\}$. Each stock $S_i$ is associated with a
discrete probability distribution $\CS_i$ over $\Delta$, where
$\CS_i(\beta)$ is the probability that the stock's return is
$\beta\%$.  For the sake of technical convenience, we allow $m_1$ and
$m_2$ to be negative.  The probability distributions
$\CS_1,\ldots,\CS_k$ are part of the input in our problem and are
obtainable, e.g., by observing historical market data.  We assume that
non-zero values satisfy $\CS_1(\beta)\geq 1/n^c$ for some constant
$c$, and when representation is important we assume that these values
can be represented as fixed-point numbers with $O(\log n)$ bits.  The
parameters $\mu$, $m_1$, and $m_2$ control the precision and range of
such observations.  For instance, for $\mu = 1$, $m_1 = 0$, and $m_2 =
200$, the set of possible returns are $0\%,1\%,\ldots,200\%$.  The
joint distribution of the $k$ probability distributions $\CS_i$ is
usually unavailable for a variety of practical reasons. In particular,
a joint distribution consists of $n^k$ entries and thus would require
observing an exponential number of data points in $k$.

\newcommand{\RAa}{{\cal RA_{\rm a}}}
\newcommand{\RAb}{{\cal RA_{\rm b}}}
\newcommand{\RAw}{{\cal RA_{\rm w}}}
\newcommand{\AGa}{{\cal AG_{\rm a}}}
\newcommand{\AGb}{{\cal AG_{\rm b}}}
\newcommand{\AGw}{{\cal AG_{\rm w}}}

\newcommand{\RAas}{{\cal RA^\ddd_{\rm a}}}
\newcommand{\RAbs}{{\cal RA^\ddd_{\rm b}}}
\newcommand{\RAws}{{\cal RA^\ddd_{\rm w}}}
\newcommand{\AGas}{{\cal AG^\ddd_{\rm a}}}
\newcommand{\AGbs}{{\cal AG^\ddd_{\rm b}}}
\newcommand{\AGws}{{\cal AG^\ddd_{\rm w}}}

The investor's goal is to find a portfolio $\vec{x}$, which is optimal
according to her risk preference in six basic cases as follows.  For a
{\it risk-averse} investor, minimizing loss is more important than
maximizing win, while an {\it aggressive} investor has the opposite
priority. Each of these two investor types can be further classified
into three subtypes, namely, {\it best-case, worst-case, and
average-case}, referring to whether the probability of loss or win is
estimated in the best, worst, or average case over the feasible joint
distributions.  More precisely, for each of these six types, the
investor first chooses a {\it target} return $\alpha$ and then looks
for such a portfolio $\vec{x}$ that optimizes one of the following six
probabilities:
\begin{itemize} 
\item $\RAb(\alpha,\vec{x})$ (respectively,
$\RAw(\alpha,\vec{x})$ or $\RAa(\alpha,\vec{x})$)
is the smallest (respectively, largest or average) probability
that the return of $\vec{x}$ is at most $\alpha\%$ over all joint
distributions for $\CS_1,\ldots,\CS_k$.
\item
$\AGb(\alpha,\vec{x})$ (respectively,
$\AGw(\alpha,\vec{x})$ or $\AGa(\alpha,\vec{x})$)
is the largest (respectively, smallest or average) probability
that the return of $\vec{x}$ is at least $\alpha\%$ over all joint
distributions for $\CS_1,\ldots,\CS_k$.
\end{itemize} 
If the investor is best-case (respectively, worst-case or
average-case) risk-averse, she would choose $\vec{x}$ to minimize
$\RAb(\alpha,\vec{x})$ (respectively, $\RAw(\alpha,\vec{x})$ or
$\RAa(\alpha,\vec{x})$).  In contrast, if the investor is
best-case (respectively, worst-case or average-case) aggressive,
she would choose $\vec{x}$ to maximize $\AGb(\alpha,\vec{x})$
(respectively, $\AGw(\alpha,\vec{x})$ or $\AGa(\alpha,\vec{x})$).

While the risk profile problem originates from a very applied field,
the corresponding mathematical model has a substantial combinatorial
structure.  In the cases where the investor is highly risk-averse or
highly aggressive, we can model the problem as a network flow problem.
Quite surprisingly, in the two-stock case, this flow problem is
solvable by a simple greedy algorithm in $O(m)$ time. In contrast, for
the three-stock case, the applicability of a greedy flow-based
algorithm would imply $P = NP$. If the number $k$ of stocks is part of
the input, we give an exact algorithm based on linear programming
which takes time polynomial in the number of entries of a
corresponding contingency table but exponential in the input size.  To
supplement this algorithm, we also give a polynomial-time
approximation algorithm based on linear programming.  We further
present an exact polynomial-time algorithm in the practical case where
the capital can only be broken up into a fixed number of units (e.g.,
cents).

It remains open whether this problem is $NP$-complete if the number of
stocks is part of the input. We strongly suspect that this is indeed
the case.

In the case of an average-case investor we show $\sharp P$-hardness of
the problem of computing the distribution function over various
probability bounds, a natural first-step in solving the average-case
investor problem.  This hardness result holds even in two dimensions,
and we describe an approximation algorithm for this case. This
algorithm uses a random walk approach to sample from the feasible
joint distributions, and is closely related to volume computation and
sampling from log-concave distributions.

Section \ref{sec_not} defines some notation. Section \ref{sec_two}
discusses the case where there are only two stocks under
consideration.  Section \ref{sec_k} discusses the case of general $k$.
Due to page limitations, all figures are placed in the appendix (these
figures are helpful in understanding the material, but are not
strictly necessary).

\section{Notation}\label{sec_not}
Let $\vec{\delta} \in \Delta^k$ denote a vector
$\port{\delta_1,\ldots,\delta_k}$, where $\delta_i \in \Delta$.  Let
\[
M=[M_{\vec{\delta}}]_{\vec{\delta}\in\Delta^k}
\]
denote a $k$-dimensional matrix indexed by $\Delta^k$.  Let $\MM_k$
denote the set of $k$-dimensional matrices for all possible joint
distributions of $\CS_1,\ldots,\CS_k$; i.e., $\MM_k$ consists of all
matrices
\[
M=[M_{\vec{\delta}}]_{\vec{\delta}\in\Delta^k},
\]
where (1) $M_{\vec{\delta}}$ is the probability that the return of
stock $S_i$ is $\delta_i\%$ for $i = 1,\ldots,k$, and (2) thus for all
$\vec{\delta} \in \Delta^k, M_{\vec{\delta}} \geq 0$ and for all
$\beta \in \Delta$ and $j = 1,\ldots,k$,
\[
\CS_j(\beta) = \sum_{\vec{\delta} \in \Delta^k; \delta_j=\beta}
M_{\vec{\delta}}.
\]
For instance, $\MM_k$ contains the matrix $M$ defined by
\[
M_{\vec{\delta}}=\prod_{i=1}^k \CS_i(\delta_i).
\]
Also, in the two-stock case, each $M \in \MM_2$ is just a
two-dimensional $m\times m$ matrix, where for all $\delta_1,\delta_2\in
\Delta$, the entries of $M$ in column $\delta_1$ sum up to
$\CS_1(\delta_1)$ and those in row $\delta_2$ sum up to
$\CS_2(\delta_2)$.

Given a portfolio $\vec{x} \in \PP_k$ and a target return $\alpha$,
let
\begin{eqnarray*}
L(\alpha,\vec{x}) & = & 
\left\{\vec{\delta}\in\Delta^k|\sum_{i=1}^k x_i\delta_i\leq\alpha\right\},
\\
L^\ddd(\alpha,\vec{x}) & = & 
\left\{\vec{\delta}\in\Delta^k|\sum_{i=1}^k x_i\delta_i < \alpha\right\},
\\
U(\alpha,\vec{x}) & = &
\left\{\vec{\delta}\in\Delta^k|\sum_{i=1}^k x_i\delta_i\geq\alpha\right\},
\\
U^\ddd(\alpha,\vec{x}) & = &
\left\{\vec{\delta}\in\Delta^k|\sum_{i=1}^k x_i\delta_i > \alpha\right\},
\end{eqnarray*}
which are the sets of the indices of all entries in the matrices in
$\Mk$ such that the return of $\vec{x}$ is at most, less than, at
least, and more than $\alpha$\%, respectively.  We further define the
following functions on $M\in\Mk$:
\begin{eqnarray*}
\hhh{L}_{\alpha,\vec{x}}(M) & = & \sum_{\vec{\delta} \in
L(\alpha,\vec{x})}M_{\vec{\delta}},
\\
\hhh{L}^\ddd_{\alpha,\vec{x}}(M) & = & \sum_{\vec{\delta} \in
L^\ddd(\alpha,\vec{x})}M_{\vec{\delta}},
\\
\hhh{U}_{\alpha,\vec{x}}(M) & = & \sum_{\vec{\delta} \in
U(\alpha,\vec{x})}M_{\vec{\delta}},
\\
\hhh{U}^\ddd_{\alpha,\vec{x}}(M) & = & \sum_{\vec{\delta} \in
U^\ddd(\alpha,\vec{x})}M_{\vec{\delta}},
\end{eqnarray*}
which are the probabilities in the joint distribution $M$ that the
return of $\vec{x}$ is at most, less than, at least, and more than
$\alpha\%$, respectively.  Formally, if $u_{\Mk}(M)$ is a uniform
density over $\Mk$,
\begin{eqnarray}
\label{p_rab}
\RAb(\alpha,\vec{x}) &=& \min_{M \in \Mk} \hhh{L}_{\alpha,\vec{x}}(M);
\\
\label{p_raw}
\RAw(\alpha,\vec{x}) &=& \max_{M \in \Mk} \hhh{L}_{\alpha,\vec{x}}(M);
\\
\label{p_raa}
\RAa(\alpha,\vec{x}) &=& \int_{\Mk}\hhh{L}_{\alpha,\vec{x}}(M)u_{\Mk}(M)dM;
\\
\label{p_agb}
\AGb(\alpha,\vec{x}) &=& \max_{M \in \Mk} \hhh{U}_{\alpha,\vec{x}}(M);
\\
\label{p_agw}
\AGw(\alpha,\vec{x}) &=& \min_{M \in \Mk} \hhh{U}_{\alpha,\vec{x}}(M);
\\
\label{p_aga}
\AGa(\alpha,\vec{x}) &=& \int_{\Mk}\hhh{U}_{\alpha,\vec{x}}(M)u_{\Mk}(M)dM.
\end{eqnarray}
For example, in the two-stock case,
$L(\alpha,\port{x_1,x_2})$ is the set of all indices in a
two-dimensional table $M$ in $\MM_2$ on or below the line $x_1
\delta_1 + x_2 \delta_2 = \alpha$, and
$\RAw(\alpha,\port{x_1,x_2})$ maximizes the sum of the
entries in this region under the condition that $M$ has the given
column and row sums of
$\CS_1(m_1),\ldots,\CS_1(m_2),\CS_2(m_1),\ldots,\CS_2(m_2)$.

For technical convenience, we also define the following terms:
\begin{eqnarray}
\label{pp_rab}
\RAbs(\alpha,\vec{x}) &=& \min_{M \in \Mk} \hhh{L}^\ddd_{\alpha,\vec{x}}(M);
\\
\label{pp_raw}
\RAws(\alpha,\vec{x}) &=& \max_{M \in \Mk} \hhh{L}^\ddd_{\alpha,\vec{x}}(M);
\\
\label{pp_raa}
\RAas(\alpha,\vec{x}) &=& \int_{\Mk}\hhh{L}^\ddd_{\alpha,\vec{x}}(M)dM;
\\
\label{pp_agb}
\AGbs(\alpha,\vec{x}) &=& \max_{M \in \Mk} \hhh{U}^\ddd_{\alpha,\vec{x}}(M);
\\
\label{pp_agw}
\AGws(\alpha,\vec{x}) &=& \min_{M \in \Mk}
\hhh{U}^\ddd_{\alpha,\vec{x}}(M);
\\
\label{pp_aga}
\AGas(\alpha,\vec{x}) &=&
\int_{\Mk}\hhh{U}^\ddd_{\alpha,\vec{x}}(M)dM.
\end{eqnarray}

\begin{lemma} \label{lem_simpl}
The following statements hold.
\begin{eqnarray}
\min_{\vec{x} \in \PP_k} \RAb(\alpha,\vec{x}) & = & 1 - \max_{\vec{x} \in \PP_k}
\AGbs(\alpha,\vec{x})
\\
\min_{\vec{x} \in \PP_k} \RAw(\alpha,\vec{x}) & = & 1 - \max_{\vec{x} \in \PP_k}
\AGws(\alpha,\vec{x})
\\
\min_{\vec{x} \in \PP_k} \RAa(\alpha,\vec{x}) & = & 1 - \max_{\vec{x} \in \PP_k}
\AGas(\alpha,\vec{x})
\\
\max_{\vec{x} \in \PP_k} \AGb(\alpha,\vec{x}) & = & 1 - \min_{\vec{x} \in \PP_k}
\RAbs(\alpha,\vec{x})
\\
\max_{\vec{x} \in \PP_k} \AGw(\alpha,\vec{x}) & = & 1 - \min_{\vec{x} \in \PP_k}
\RAws(\alpha,\vec{x})
\\
\max_{\vec{x} \in \PP_k} \AGa(\alpha,\vec{x}) & = & 1 - \min_{\vec{x} \in \PP_k}
\RAas(\alpha,\vec{x})
\end{eqnarray}
\end{lemma}
\begin{proof}
Straightforward.
\end{proof}

In light of Lemma~\ref{lem_simpl}, to solve the risk profile problem,
it suffices to show how to compute
\[
\begin{array}{lll}
\min_{\vec{x} \in \PP_k} \RAb(\alpha,\vec{x}), & 
\min_{\vec{x} \in \PP_k} \RAw(\alpha,\vec{x}), &
\min_{\vec{x} \in \PP_k} \RAa(\alpha,\vec{x}),
\\
\min_{\vec{x} \in \PP_k} \RAbs(\alpha,\vec{x}), &
\min_{\vec{x} \in \PP_k} \RAws(\alpha,\vec{x}), &
\min_{\vec{x} \in \PP_k} \RAas(\alpha,\vec{x}).
\end{array}
\]
The techniques for 
computing the latter three expressions are essentially the same as those for 
computing the former three. Furthermore,
the techniques for computing the first expression are  
almost identical to those for computing the second.
For these reasons, the remainder of our discussion focuses on how to compute 
$\min_{\vec{x} \in \PP_k} \RAw(\alpha,\vec{x})$ and $\min_{\vec{x} \in
\PP_k} \RAa(\alpha,\vec{x})$.

\section{The Two-Stock Case}\label{sec_two} 
\begin{figure}[h]
\begin{center}
\epsfig{figure=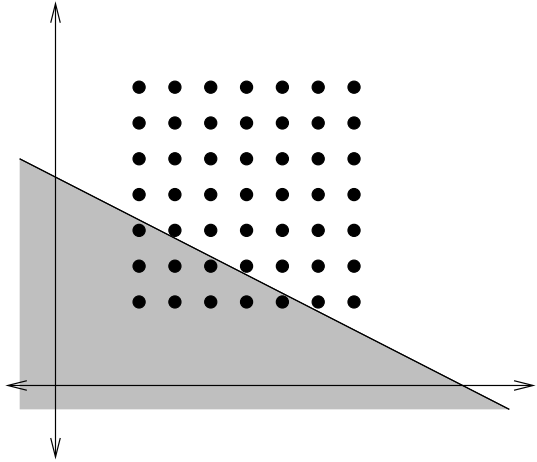}
\end{center}
\caption{Visualization of two stock case}
\label{fig:visual}
\end{figure}

This section assumes that $k=2$, i.e., there are only two stocks under
consideration.  In the case of two stocks, we can visualize the
problems under consideration as in Figure~\ref{fig:visual}.  The
discrete and finite set of possible return pairs for the two stocks in
the portfolio are shown as the dots in this picture -- each pair has a
probability (from the joint distribution) associated with it, with the
given restrictions on column and row sums.  A given portfolio and
target return $\alpha$ defines a half-space on the set of return
pairs, with the shaded area in Figure~\ref{fig:visual} giving the area
in which the total return is $\leq\alpha$.  The problem of computing
$\RAw(\alpha,\vec{x})$ then is the problem of determining which
feasible assignment of joint probabilities places the highest total
probability in the shaded region.

\subsection{A Worst-Case or Best-Case Investor} 
Given a target return $\alpha$, this section focuses on how to compute
an optimal portfolio for a worst-case risk-averse investor. The cases
of a best-case risk-averse investor, a worst-case aggressive investor,
and a best-case aggressive investor can be solved similarly.

We first present a basic algorithm to compute $\RAw(\alpha,\vec{x})$
by computing a worst-case joint distribution matrix $M$ for $\CS_1$
and $\CS_2$. For convenience, we index the entries of $M$ with
$\{(i,j) \mid i, j = m_1, \ldots, m_2\}$, where row $i$ (respectively,
column $i$) corresponds to return $i\mu$ of $S_1$ (respectively,
$j\mu$ of $S_2$).  We model the problem of computing $M$ as a network
flow problem on the graph $G$ defined below:
\begin{itemize}
\item 
\sloppy
$G$ has $2(m+1)$ vertices, namely, a source $s$, a sink $t$, and
$v_{m_1},\ldots,v_{m_2}$, $w_{m_1},\ldots,w_{m_2}$, where $v_i$
(respectively, $w_i$) corresponds to return $i\mu$ of stock $S_1$
(respectively, stock $S_2$).

\item 
For all $i, j = m_1,\ldots,m_2$, $G$ has (1) edge $(v_i,w_j)$, which
has capacity $c(v_i,w_j)=1$ if $x_1i\mu+x_2j\mu\leq \alpha$ or $0$
otherwise; (2) the edge $(s,v_i)$ with capacity
$c(s,v_i)=\CS_1(i\mu)$; and (3) the edge $(w_j,t)$ with capacity
$c(w_j,t)=\CS_2(j\mu)$.
\end{itemize}

Geometrically, we wish to push as much probability as possible into
the region of $M$ defined by $x_1 i + x_2 j \leq \frac{\alpha}{\mu}$.
In other words, the value of a maximum $s-t$ flow of $G$ equals
$\RAw(\alpha,\vec{x})$.  Thus, it is tempting to use a maximum flow
algorithm to solve this maximum flow problem.  The fastest known
algorithm for this problem is due to Goldberg and Rao \cite{Goldberg:1998:BFD} and
runs in $O^*(m^{2\frac{2}{3}})$ time\footnote{We use $O^*(f(n))$ for
the ``soft-O'' notation, which ignores polylogarithmic factors.  In
bounds for the approximation algorithms, this notation also ignores
factors that depend only on the approximation bound $\epsilon$.} for
our application (note that $m$ in this bound is as defined in this
work, not as the number of edges which is typical in general flow
discussion).  Instead of using this algorithm, we exploit some
structural properties of $G$ to solve the flow problem using a simple
greedy algorithm in $O(m)$ arithmetic operations.  Note that since $G$
may have $\Omega(m^2)$ edges with positive capacity, we cannot afford
to construct the whole $G$ explicitly. The idea of our $O(m)$-time
algorithm can be described as follows.

Starting with $v_{m_2}$, we try to push a flow of $c(s,v_{m_2})$
through $G$. Assume $c(v_{m_2},w_{m_1})=1$ for simplicity.  We
consider the path formed by edges
$(s,v_{m_2}),(v_{m_2},w_{m_1}),(w_{m_1},t)$ first.  We can push flow
\sloppy $\min(c(s,v_{m_2}),c(w_{m_1},t))$ through this path, saturating either
$(s,v_{m_2})$ or $(w_{m_1},t)$.  If we saturated $(s,v_{m_2})$ then we
next consider the path $(s,v_{m_2-1})$, $(v_{m_2-1},w_{m_1})$,
$(w_{m_1},t)$ for pushing additional flow; however, if we had
saturated $(w_{m_1},t)$ we will next consider the path
$(s,v_{m_2}),(v_{m_2},w_{m_1+1}),(w_{m_1+1},t)$.  We continue in this
fashion until we can push no more flow.  The only complication is that
if at some point we are considering the path
$(s,v_i),(v_i,w_j),(w_j,t)$, and $c(v_i,w_j)=0$, then obviously we
can't saturate \emph{either} $(s,v_i)$ or $(v_j,t)$, and we simply
decrease $i$ to next consider the path
$(s,v_{i-1}),(v_{i-1},w_j),(w_j,t)$.  The details of this $O(m)$ time
algorithm are given in Figure~\ref{greedy}.

\newcommand{\Greedy}{{\rm  Greedy-Flow}}

\newcounter{greedy}
\begin{figure}[h]
\begin{algorithmic}
\STATE \textbf{procedure} Greedy-Flow
\STATE
\STATE $F\leftarrow 0$
\STATE $i\leftarrow m_2$
\STATE $cv\leftarrow c(s,v_i)$
\STATE $j\leftarrow m_1$
\STATE $cw\leftarrow c(w_j,t)$
\LOOP
\IF{$c(v_i,w_j)=1$ \textbf{and} $cw\leq cv$}
\STATE $F\leftarrow F+cw$
\STATE $cv\leftarrow cv-cw$
\STATE $j\leftarrow j+1$
\STATE \textbf{if } $j>m_2$ \textbf{then return} $F$
\STATE $cw\leftarrow c(w_j,t)$ 
\ELSE
\IF{$c(v_i,w_j)=1$}
\STATE $F\leftarrow F+cv$
\STATE $cw\leftarrow cw-cv$
\ENDIF
\STATE $i\leftarrow i-1$
\STATE \textbf{if } $i<m_1$ \textbf{then return} $F$
\STATE $cv\leftarrow c(s,v_i)$
\ENDIF
\ENDLOOP
\end{algorithmic}
\caption{The procedure \Greedy}\label{greedy}
\end{figure}

\begin{theorem}
Given $\CS_1, \CS_2$, a valid portfolio vector $\vec{x}$, and $\alpha$
as input, \Greedy\ computes the value of a maximum flow of $G$ in
$O(m)$ arithmetic operations.
\end{theorem}
\begin{proof}
As a first step we prove that the algorithm computes the maximal
flow. Let $\ell$ be the minimal index such that $(w_\ell,t)$ is not
saturated after termination of the algorithm and $k$ be the minimal
index such that $c(v_k,w_\ell)=0$.  We define a partition $V_1\cup V_2$
of the nodes by
\[
V_1=\{ s,v_k,\ldots,v_{m_2},w_{m_1},\ldots,w_{\ell-1}\},
\quad V_2=\bar{V_1}.
\]
It is trivial from the definition of $j$ that the edges $e=(w_i,t),
i=\{m_1,\ldots,\ell-1\}$ are saturated.

Since $x_1, x_2 \geq 0$, and $k$ is the minimal value such that
$c(v_k,w_\ell)=0$, we have $c(v_i,w_\ell)=1$ for
$i=m_1,\ldots,k-1$. Since $(w_\ell,t)$ is not saturated, all edges
$(s,v_i), i\in\{m_1,\ldots,k-1\}$ must be saturated.

From the definition of $k$ and the non-negativity of the portfolio
vector it is easy to see that edges $e=(v_i,w_j)$ for
$i\in\{k,\ldots,m_2\}$, $j\in\{\ell,\ldots,m_2\}$ and positive
capacity cannot exist.  Thus, every edge $e=(x,y)$ with $x\in V_1$ and
$y\in V_2$ is saturated.  The Max-Flow-Min-Cut Theorem then implies
that the algorithm indeed computes a maximal flow.

Observing the fact that in each loop iteration either index $i$ is
decremented or index $j$ is incremented, and that there are only $m$
different values that either $i$ or $j$ can take on before the
algorithm terminates, there are at most $2m-1$ loop iterations, and
the linear running time bound follows.
\end{proof}

To compute $\inf \{\RAw(\alpha,\vec{x})|\sum x_i=1\}$ we have to compute
$\RAw(\alpha,\vec{x})$ for all possible portfolios $\port{x_1,x_2}$.
However, each feasible portfolio corresponds to a half-space (as
in Figure~\ref{fig:visual}) defined by a line that goes through the
point $(\alpha,\alpha)$ ($x_1\alpha+x_2\alpha=\alpha$, since
$x_1+x_2=1$), so we only need to consider the $O(m^2)$ distinct subsets
of return pairs that can be defined by a line going through
$(\alpha,\alpha)$.  We can identify each such portfolio with
a different (non-positive) slope $s_1, \ldots, s_{m^2}$, which we
assume to be sorted in descending order. By using a suitable data
structure it is possible to compute the best portfolio much faster
than the obvious $O(m^3)$ algorithm that starts the greedy algorithm
for each slope.

\begin{theorem}
Given $\CS_1, \CS_2$, and $\alpha$, we can compute in $O(m^2\log m)$
arithmetic operations a portfolio $\port{x_1,x_2}$ for a worst-case
risk-averse investor which minimizes equation~$($\ref{p_raw}$)$.
\end{theorem}
\begin{proof}
Starting with the first slope $s_1$ we build up a binary tree. Each is
labeled with a pair of two real entries $(e_1,e_2)$. The leaves of the
tree correspond to the rows and the columns in the following way.

Starting from column $m_2$ we add leaves from left to right.  We add
leaves with labels $(0,\CS_2(m_1\mu))$,
$(0,\CS_2((m_1+1)\mu)),\ldots$, $(0,\CS_2(j_{m}\mu))$, until we reach
a row index $j_{m}$ such that $x_1m_2\mu+x_2(j_m+1)\mu>\alpha$, i.e.,
this index is the last under the crucial line.  To be precise we let
$j_m=\lfloor\frac{\alpha-x_1m_2\mu}{x_2\mu}\rfloor$; note that it
may be the case that $j_m<m_1$, so this sequence of leaves may be
empty.  Then we add the leaf $(-\CS_1(m_2\mu), 0)$.  Next, we consider
column $m_2-1$ and add leaves $(0,\CS_2((j_m+1)\mu), \ldots,
(0,\CS_2((j_{m-1})\mu))$, until we reach an index $j_{m-1}$, such that
$x_1(m_2-1)\mu+x_2(j_{m-1}+1)\mu>\alpha$. Then we add the leaf
$(-\CS_1((m_2-1)\mu), 0)$ and proceed similarly with column $m_2-2$.
Note that the order of adding leaves is crucial to this data
structure and the correctness of the algorithm is based on
that. Starting from left to right we group the leaves in pairs of 2
and build a parent node for each pair according to the following rule
\[
\mbox{parent}[(e_1,e_2),(f_1,f_2)]=
(e_1+\min\{e_2+f_1,0\},\max\{e_2+f_1,0\}+f_2).
\]
We build $O(\log m)$ layers iteratively, until we reach a single root
node $(r_1,r_2)$.  It is easy to see that this tree based algorithm
imitates the greedy algorithm described before and that $1+r_1=1-r_2$
is exactly the flow value.  Building this tree structure takes
constant time per tree node, and since there are $O(m)$ nodes we have
a total time of $O(m)$, which is no better than the time bound of the
greedy algorithm. The advantage is that we can dynamically update this
data structure efficiently.

We will first sort all of the $m^2$ possible return pairs by their
slope with the point $(\alpha,\alpha)$, so that as the slope
determined by our portfolio increases we can quickly (in constant time
per pair) determine which pairs are added and which are removed from
our half-space of interest.  This takes $O(m^2\log m)$ time.  To update
our data structure for each point insertion/removal, all that is
required is swapping the position of two neighboring leaves.  With
obvious techniques, the positions of these two leaves can be found in
$O(1)$ time, and we can update the tree by looking at the path from the
two leaves to the root and update each node on that path. Each update
step requires $O(1)$ operations and the length of the path is bounded
by $O(\log m)$.  Since there are at most $m^2$ point additions and
removals, each taking $O(\log m)$ time, it takes at most $O(m^2\log
m)$ time to consider all possible portfolios.
\end{proof}

\subsection{The Average-Case Investor}
For the average-case investor ($\RAa$ or $\AGa$), we are not
interested in the extremes of the joint distributions, but rather the
distribution of the feasible tables.  In this section we consider
$Q=\hhh{L}_{\alpha,\vec{x}}(M)$ a random variable where $M$ is drawn
from a uniform distribution over the feasible tables $\MM_k$.  The
definition of $\RAa(\alpha,\vec{x})$, from~(\ref{p_raa}), is then
$E[Q]$.  We will see that computing the distribution function of $Q$
is a computationally difficult problem to solve exactly, but can be
approximated within a reasonable (polynomial) amount of time.

\begin{theorem}
Let $\gamma\in [0,1]$ be an $n$-bit rational.  It is $\sharp P$-hard
to compute the fraction of feasible tables $M\in \MM_2$ with
\[
\hhh{L}_{\alpha,\vec{x}}(M)=
  \sum_{\delta \in L(\alpha,\vec{x})} M _{\delta}\leq \gamma
\]
$($the integration of the corresponding indicator function, or the
distribution function for $Q)$. 
\end{theorem} 
\begin{proof}
Given positive integers $a_1,\ldots,a_n,b$, it is shown in
\cite{Dyer:1991:CVC} that computing the $n$-dimensional volume of the
polyhedron $P$
\[
\sum_{j=1}^n a_j y_j\leq b\qquad 0\leq y_j\leq 1\quad(j=1,\ldots,n)
\]
is $\sharp P$-hard.  Let $d=\sum_{j=1}^{n} a_j$ and consider the
polyhedron
\begin{equation}
\sum_{j=1}^{n+1} a_j y_j = d\qquad 0\leq y_j\leq 1\quad(j=1,\ldots,n+1),
\label{eq:np1sum}
\end{equation}
where $a_{n+1}=d$.  Note that for any valid assignment of values to
$y_1,y_2,\ldots,y_n$ we have $0\leq\sum_{j=1}^{n} a_jy_j\leq d$, so
there is a $y_{n+1}\in[0,1]$ that will satisfy~(\ref{eq:np1sum}).  Now
let $a'_i=a_i/(2d)$ and define a $2\times (n+1)$ contingency table by
$t_{1j}=a'_jy_j, t_{2j}=a'_j(1-y_j)$, with row sums $(1/2,1/2)$ and
column sums $(a'_1,\ldots,a'_{n+1})$.

To completely define our stock problem, we must also give values for
$\mu$, $\alpha$, the portfolio $\vec{x}=\port{x_1,x_2}$, and the
threshold $\gamma$, which we
do as follows:
\[
  \mu=1 , \hspace*{0.3in} x_1 = \frac{1}{n+1} , \hspace*{0.3in} 
   x_2 = \frac{n}{n+1} , \hspace*{0.3in} \alpha=\frac{2n}{n+1} ,
   \hspace*{0.3in} \gamma=\frac{b}{2d} .
\]
It is straightforward to verify from these values that the return
pairs in the critical region (the shaded region in
Figure~\ref{fig:visual}) are exactly the entries
$t_{1j}$ for $j=1,\ldots,n$.  Therefore, the tables that satisfy our
criteria, that $\hhh{L}_{\alpha,\vec{x}}(M)\leq\gamma$, are precisely
those with
\[
   \sum_{j=1}^n t_{1j} \leq\gamma 
      \hspace*{0.2in}\Longleftrightarrow\hspace*{0.2in}
   \sum_{j=1}^n a'_jy_j \leq\gamma
      \hspace*{0.2in}\Longleftrightarrow\hspace*{0.2in}
   \sum_{j=1}^n a_jy_j \leq \gamma \cdot 2d = b .
\]
Therefore the feasible tables that meet our criteria are exactly those
that correspond to points in polyhedron $P$, and so the fraction of
tables that meet the criteria is exactly the volume of $P$.
\end{proof}

Following the notation of Dyer, Kannan and Mount \cite{Dyer:1997:SCT}, who
describe a sampling procedure for contingency tables with integer
entries and large row and column sums ($\geq\Omega(m^3)$), we define
\begin{eqnarray*}
V(r,c) & = & \left\{x\in\R^{m\times m}|\sum_jx_{ij}=r_i\ \mbox{for}\ i=
1,\ldots,m\ \right.
\\
& & \left.\mbox{and}\ \sum_ix_{ij}=c_j \mbox{ for }
j=1,\ldots,m\right\}
\end{eqnarray*}
and 
\[
P(r,c)=V(r,c)\cap\{x|x_{ij}\geq 0 \mbox{ for } i= 1,\ldots,m,
j=1,\ldots,m\}
\] 
as the contingency polytope.  Thus, $V(r,c)$ is the set of matrices
with row and column sums specified by $r$ and $c$ respectively. In our
case $r_i=\CS_1(i\mu)$, $c_i=\CS_2(i\mu)$ and $P(r,c)$ is the set of joint
distributions $\Mk$.

Let $U$ be the lattice 
\[
\{x\in\Z^{m\times m}|\sum_jx_{ij}=0 \mbox{ for } i=
1,\ldots,m,\sum_ix_{ij}=0 \mbox{ for } j=1,\ldots,m\}.
\] 
For $1\leq i\leq m-1$ and $1\leq j\leq m-1$, let $b(ij)$ be the vector
in $\R^{m\times m}$ given by $b(ij)_{i,j}=1,
b(ij)_{i+1,j}=-1,b(ij)_{i,j+1}=-1,b(ij)_{i+1,j+1}=1$ and
$b(ij)_{k,\ell}=0$ for all other indices $k,\ell$.  Any vector $x$ in
$V(0,0)$ can be expressed as linear combination of the $b(ij)$'s as
follows 
\[
x=\sum_{k=1}^{m-1}\sum_{\ell=1}^{m-1}\left(\sum_{i=1}^{k}
\sum_{j=1}^{\ell}x_{ij}\right)b(k\ell).  
\] 
It is easy to see that the $b(ij)$ are all linearly independent and
the the dimension of $V(r,c)$ and $P(r,c)$ for positive row and column
sum vectors $r$ and $c$ is $(m-1)^2$ \cite{Dyer:1997:SCT}.  We will apply the
sampling algorithm pioneered by Dyer, Frieze and Kannan \cite{Dyer:1991:RPT}
and later refined in a sequence of papers (see \cite{Kannan:1994:MCP} for an
overview) to sample uniformly at random in $P(r,c)$.

We sample in the space $V(r,c)$. As mentioned in the introduction, we
know a starting point $z_0$ in $P(r,c)$ (multiplication of rows and
column sums).  It is easy to see that a ball of radius $b^2$ is
inside $P(r,c)$, if every component of $r$ and $c$ is at least $b$.
Since in our case $r$ and $c$ sum up to one, $P(r,c)\subset B(0,1)$.
The following theorem is a corollary of the analysis of the fastest
sampling algorithm in convex bodies known so far by Kannan, Lov{\'a}sz and
Simonovits \cite{Kannan:1997:RWV}.

\begin{theorem}\label{sampling} We can generate a point in $P(r,s)$,
which is almost uniform in the sense that its distribution is at most
$\epsilon$ away from the uniform in total variation distance.  The
algorithm uses $O^*(\frac{m^6}{b^4})$ membership queries of $P(r,s)$
(each requires $O(m^2)$ arithmetic operations).  
\end{theorem}

\newcounter{estimate}
\begin{figure}[h]
\begin{algorithmic}
\STATE \textbf{procedure} Estimate$(x)$
\STATE
\STATE $S\leftarrow 0$
\STATE $N=\frac{100}{\epsilon^2\delta}$
\FOR{$\ell=1,\ldots,N$}
\STATE $\zeta_i\leftarrow$ result from sample procedure started at $x$
\STATE $S\leftarrow S+\hhh{L}_{\alpha,\vec{x}}(\zeta_i)$
\ENDFOR
\STATE $S\leftarrow S/N$
\STATE \textbf{return} $S$
\end{algorithmic}
\caption{The approximation algorithm}\label{approx}
\end{figure}

\begin{theorem}
Procedure Estimate $($in Figure~\ref{approx}$)$ computes a number $S$
in $O^*\left(\frac{m^8}{b^4\epsilon^2\delta}\right)$ arithmetic
operations, which approximates $\RAa(\alpha,\vec{x})$ $($i.e.,
$\RAa(\alpha,\vec{x})-\epsilon\leq S\leq
\RAa(\alpha,\vec{x})+\epsilon)$ with probability $1-\delta$.
\end{theorem}
\begin{proof}
Let $S_k=\frac{1}{k}\sum_{i=1}^k
\hhh{L}_{\alpha,\vec{x}}(\zeta_i)$. Thus, $E(S_k)= \int
\hhh{L}_{\alpha,\vec{x}}(M)w(M)dM$, where $w$ is the density
produced by the random walk. Since $0\leq\hhh{L}_{\alpha,\vec{x}}(M)\leq 1$
for all $M\in\MM_2$, it is easy to see that $\sigma^2(S_1)\leq 1$ and so
$\sigma^2(S_k)\leq \frac{1}{k}$. By Chebychev's inequality,
\[
P(|S_k-E(S_k)|\geq\epsilon/2)\leq \frac{\sigma^2(S_k)}{(\epsilon/2)^2}\leq
\frac{4}{\epsilon^2 k} .
\]
Since the samples are not entirely uniform, we must consider the error
introduced by the approximately uniform sampling distribution as well.
Let $u_{\Mk}(M)$ denote a uniform density over the set
$\MM_k$, and then approximating a uniform distribution within bound
$\epsilon/4$, Theorem~\ref{sampling} implies
\begin{eqnarray*}
\hbox to 0.2in{$|E(S_k)-\RAa(\alpha,\vec{x})|$\hss}\\
&=&
\left|\int \hhh{L}_{\alpha,\vec{x}}(M)w(M)dM-
  \int \hhh{L}_{\alpha,\vec{x}}(M)u_{\Mk}(M)dM\right|\\
&\leq&\int_{w>u_{\Mk}}
\left(w(M)-u_{\Mk}(M)\right)dM
\\ & & 
+
\int_{w\leq u_{\Mk}}
\left(u_{\Mk}(M)-w(M)\right)dM\\
&\leq& \epsilon/2 .
\end{eqnarray*}
Setting $k=\frac{4}{\epsilon^2\delta}$ the theorem follows.
\end{proof}

\section{The $k$-Stock Case}\label{sec_k}
In this chapter we consider the general case of more than two stocks.
Since the problem of estimating the probability distribution for the
average-case investor is already $\sharp$-P complete in the two stock
case, we do not consider it any more and concentrate on a
worst-case investor.  We start with a complexity result for three
stocks, which implies that a greedy or flow based portfolio is quite
unlikely to exist.

\begin{theorem}
The existence of a greedy or flow based portfolio for the problem with
3 or more stocks implies $P=NP$.
\end{theorem}
\begin{proof}
We prove this result by reduction from NUMERICAL-3-DIM-MATCHING.
Consider an instance of NUMERICAL-3-DIM-MATCHING, i.e., disjoint sets
$X_1,X_2,X_3$, each containing $m$ elements, a size $s(a)\in\Z^+$ for
each element $a\in X_1\cup X_2\cup X_3$ and bound $B\in\Z$.  We would
like to know if $X_1\cup X_2\cup X_3$ can be partitioned into $m$
disjoint sets such that each of these sets contains exactly one
element from each of $X_1$, $X_2$, and $X_3$, and the sum of the
elements is exactly $B$ (we can change this requirement to $\leq B$
without difficulty).  This problem is NP-complete in the strong sense,
so we restrict the sizes to be bounded by a polynomial, $s(a)\leq n^c$
for some constant $c$.

We construct an instance of the problem of computing
$\RAw(\alpha,\port{1/3,1/3,1/3})$ by making a contingency table in
which $\CS_k(i)=c_{k,i}/m$, where $c_{k,i}$ is the number of items in
set $X_k$ with value $i$.  The existence of a greedy or flow based
algorithm implies the existence of a solution in which all entries in
the solution table are multiples of $1/m$, and such a solution exists
with $\hhh{L}_{\alpha,\vec{x}}(M)=1$ if and only if there is a valid
partition of $X_1\cup X_2\cup X_3$.  If such a partition exists, we
can find it by simply taking all of the triples ``selected'' (with
multiplicity determined by the integer multiple of $1/m$), and use
elements from $X_1$, $X_2$, and $X_3$ as determined by the three
coordinates of each selected point.
\end{proof}

While this proof shows that it is unlikely that a fast and simple
greedy or flow-based algorithm exists, as it does for 2 stocks, we can
indeed solve the problem for a fixed number of stocks in polynomial
time using a more time-consuming procedure based on linear
programming.  This is stated in a general setting in the following
theorem.

\begin{theorem}
If the number of stocks $k$ is part of the input, the problem of
determining the best portfolio for a worst-case investor can be solved
in time polynomial in the number of entries of the contingency table
(but exponential in $k$).
\end{theorem}
\begin{proof}
The problem can be modeled as linear program with a number of
variables, that corresponds to the number of entries of the
contingency table, and $km$ inequalities.
\end{proof}

\subsection{An Approximation Algorithm}
In this section we describe an approximation algorithm, that solves
the problem of determining the worst case probability for a given
portfolio within a given error $\epsilon\in \R^+$ in polynomial time.
Additionally, we describe an important, non-trivial special case,
where the problem can be solved exactly in polynomial time.

\begin{theorem}
Suppose that a portfolio $\port{x_i}_{i=1}^k$ and a target return
$\alpha$ are given. The worst-case probability can be
approximated $($i.e., we compute a value $W$ with
$\RAw(\alpha,\vec{x})-\epsilon \leq W \leq
\RAw(\alpha,\vec{x})+\epsilon)$ in time polynomial in $k$ and $n$. The
number of steps is dominated by solving a linear program in
$O(km^2/\epsilon^2)$ variables and $O(km/\epsilon)$ constraints.
\end{theorem}
\begin{proof}
We consider the first pair of stocks $S_1$ and $S_2$ as in the two
dimensional case and define a new portfolio as
$\tilde{x_1}=\frac{x_1}{x_1+x_2}$ and
$\tilde{x_2}=\frac{x_2}{x_1+x_2}$.  We divide the two dimensional
plane in $\ell=\frac{1}{\epsilon}m\log k $ regions by $\ell$ parallel
lines $ \tilde{x_1}x+\tilde{x_2}y=const$ of constant
distance. Thus, we divide the entries of the joint distribution matrix
into $\ell$ different sets (see Figure \ref{LP}).
\begin{figure}[h]
\begin{center}
\epsfig{figure=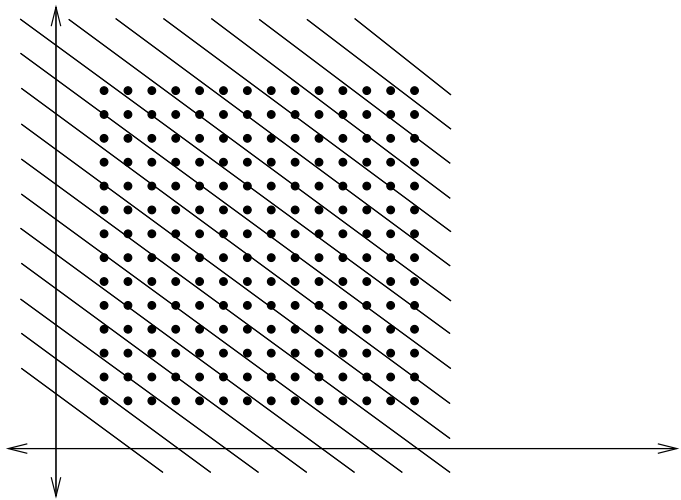}
\end{center}
\caption{Striping idea used in worst-case approximation construction}
\label{LP}
\end{figure}

Each entry in the matrix corresponds to a variable and the variables
satisfy the row sum and column sum condition of the joint
distribution. Next, we sum up the entries in the $\ell$ different sets
and assign the sums to $\ell$ new variables.  By combining these sum
variables from two different pairs of stocks, we get a new table with
new row and column sum conditions, resulting again in $\ell$ new sum
variables.

Repeating combinations in this manner, we stop after $\log k$
iterations and the creation of $O(km^2\log k/\epsilon^2)$ variables
and $O(km\log k/\epsilon)$ constraints, leaving just one table
with 2 border distributions (expressed as variables).  Assuming, that
the variables of the border distributions correspond to the
distribution of the stocks $S_1,\ldots,S_{k/2}$ and
$S_{k/2+1},\ldots,S_{k}$, we do the following.

We define a portfolio $\tilde{x_1}=\frac{x_1+\cdots+x_{k/2}}{\sum
x_i}$ and $\tilde{x_2}=\frac{x_{k/2+1}+\cdots+x_{n}}{\sum x_i}$ for
our last table and consider the line $\tilde{x_1}x+
\tilde{x_2}y=\alpha$, dividing our last table in two sets.  The
variables below that line are summed up and we solve a linear program
by maximizing this sum subject to the constraints created before.
Since we reduced the number of entries in each table from
$\Omega(m^2)$ to only $\ell$, that are considered in the next table,
we lost some precision during the combination. But, after the first
pairing in the lowest level of the binary tree, each sum variable
represents a loss probability of the combination of the two stocks
within an error of $\frac{\epsilon}{\log k}$\%. Furthermore, it is
easy to see that during the repeated combination of the stocks the
error accumulates linearly in each iteration. Thus, the theorem
follows.
\end{proof} 

\begin{theorem}
Suppose that a portfolio $\port{x_i}_{i=1}^k$ and a target return
probability $p$ is given. Under the assumption, that the dollar, that
has to be invested, can only be broken into a fixed number $c$ of
equal units (cents), the worst-case probability can be computed
exactly in time polynomial in $k$ and $m$.
\end{theorem}
\begin{proof}
The proof is based on a similar construction as the approximation
algorithm and is omitted for brevity.
\end{proof}

\section*{Acknowledgments}
The authors wish to thank the anonymous referees for very helpful
comments.  

A preliminary version of this work appeared in {\em Proceedings of the
32nd Annual {ACM} Symposium on Theory of Computing}, pages 228--234,
2000.

\bibliographystyle{apalike}
\chapbblname{kao_risk}
\chapbibliography{kao_risk}
\end{document}